\documentclass[fleqn,usenatbib]{mnras}

\usepackage{newtxtext,newtxmath}
\usepackage{times}
\usepackage[T1]{fontenc}
 \usepackage[normalem]{ulem}

\DeclareRobustCommand{\VAN}[3]{#2}
\let\VANthebibliography\thebibliography
\def\thebibliography{\DeclareRobustCommand{\VAN}[3]{##3}\VANthebibliography}

\usepackage{graphicx}	
\usepackage{amsmath}	
\usepackage{booktabs}

\newcommand{\Msun}{$h^{-1}{\rm M}_{\odot}$~}

\title[The Ultramarine Simulation]{The Ultramarine Simulation: properties of dark matter haloes before redshift 5.5}

\author[Q. Wang et al.]{
Qiao Wang,$^{1,2}$\thanks{E-mail: qwang@nao.cas.cn}
Liang Gao,$^{1,2,3,4}$
Chen Meng$^{1}$
\\
$^{1}$Key Laboratory for Computational Astrophysics, National Astronomical Observatories, Chinese Academy of Sciences, Beijing 100101, China\\ 
$^{2}$Institute for Frontiers in Astronomy and Astrophysics, 
Beijing Normal University,  Beijing 102206, China\\
$^{3}$School of Astronomy and Space Science, University of Chinese Academy of Sciences, Beijing 100049, China\\
$^{4}$Institute for Computational Cosmology, Department of Physics, Durham University, Science Laboratories, South Road, Durham DH1 3LE, England\\
}

\date{Accepted XXX. Received YYY; in original form ZZZ}

\pubyear{2022}

\begin{document}
\label{firstpage}
\pagerange{\pageref{firstpage}--\pageref{lastpage}}
\maketitle

\begin{abstract}
We introduce the Ultramarine simulation, an extremely large $N$-body simulation of the structure formation and evolution to redshift 5.5 at which cosmic reionization was just completed. The simulation evolves 2.1 trillion particles within a $512~h^{-1}$Mpc cube and has an unprecedented mass and force resolution for large volume simulations of this kind, $5.6\times 10^6 ~h^{-1}$M$_{\odot}$ and 1.2 $h^{-1}$kpc, respectively. We present some basic statistical results of the simulation, including the halo mass function, halo bias parameter as well as halo mass-concentration relation at high redshifts, and compare them with some existing representative models.  We find excellent agreement with some models on the high redshift halo mass functions, but neither the halo bias factor nor halo mass-concentration relation. All halo bias models for comparison over-predicate high redshift halo bias by large factors, an accurate fit to our simulation is given. High redshift dark matter haloes still can be reasonably described with NFW model, the halo mass-concentration relations are monotonic, with more massive haloes having lower concentration, in disfavor of the upturn feature reported by some studies. The mass concentration relation has little evolution between $z=5.5$ to $z=10$, in contrast to strong evolution predicted by most existing models. In addition, concentration parameters of high redshift dark matter haloes are much lower than most model predictions.
\end{abstract}


\section{Introduction}
Progresses on observational extra-galactic astronomy  have been made in recent decades greatly broaden our knowledge of our   universe, especially the near universe. Meanwhile, numerical simulation is essential to understand or interpret the large body of observational data because of the nonlinear nature of   cosmic structure formation and evolution. Recent cosmological simulations not only are able to predict abundance and clustering of galaxies and their dark matter haloes \citep{2005Natur.435..629S,  2008MNRAS.391.1685S, 2009MNRAS.398.1150B, 2011ApJ...740..102K, 2012MNRAS.423.3018P, 2012MNRAS.426.2046A, 2012MNRAS.425.2169G, 2015ApJS..219...34H, 2015PASJ...67...61I, 2016MNRAS.457.4340K, 2016NewA...42...49H, 2016PASJ...68...25M, 2017ComAC...4....2P, 2018ApJS..236...43G, 2019ApJS..245...16H, 2020NatRP...2...42V, 2021ApJS..252...19H, 2021MNRAS.508.4017M, 2021arXiv210901956F, 2021MNRAS.506.4210I, 2022LRCA....8....1A}, but also their internal properties, for example, morphological types, metalicity as well as some gaseous properties  \citep{2014Natur.509..177V,2015MNRAS.446..521S,2015MNRAS.452..575S, 2016A&C....15...72M}.  

With the successful launch of James Webb Space Telescope (JWST)~\citep{2006SSRv..123..485G}, along with the forthcoming  Square Kilometre Array Phase 1 (SKA1)~\citep{2015aska.confE.174B, 2019arXiv191212699B}, much attention will be focused on high redshift  universe at which there were very limited observations \citep{2015ApJ...803...34B, 2017ApJ...844...85O, 2018ApJ...855..105O,  2019MNRAS.488.4271G, 2020MNRAS.493.1662M}, in particularly, the evolution of the  universe in the first billion years. These programs will reveal when and how the first galaxies emerge and how our universe was reionized by them. Theoretically, numerical simulations involving radiative transfer processes are still the most powerful tool to understand the cosmic reionization. These simulations can roughly be classified into two categories, namely radiative hydrodynamical simulations \citep{2013ApJ...776...81B, 2014ApJ...793...29G, 2016MNRAS.463.1462O, 2019MNRAS.485..117K, 2020MNRAS.496.4087O, 2022MNRAS.512.4909G} and hydrodynamical/$N$-body simulations combined with  post-processing methods \citep{2007ApJ...671....1T, 2015ApJ...813...54T, 2019MNRAS.489.5594M, 2019MNRAS.486.4075O, 2020ApJ...905...27S}. While the first approach directly follows structure formation and instantaneously solves radiative transfer process, and thus provide most detailed description of the physical processes, it is computationally challenging. Simulations with this approach are usually constrained to be small volume. Previous studies show that ionized bubbles could extend over size of tens of Mpc \citep{2004ApJ...613....1F}, and  simulation box $\leq 100$ Mpc tends to underestimate the large-scale power and induce bias and scatter \citep{ 2013ApJ...776...81B, 2014MNRAS.439..725I, 2015aska.confE...7I, 10.1093/mnras/staa1323}. 
The second approach utilizes N-body or hydrodynamical simulation and carry out radiative transfer calculation with post-processing. As the N-body simulation is much cheaper and so can readily generate large volume density field and resolve low dark matter haloes of ionized sources. Recent examples with this approach include \citet{2006MNRAS.369.1625I} and \citet{2007MNRAS.377.1043M}.

Aiming to understand what happened in the first billion years, as a step, we perform the Ultramarine simulation, an extremely large dark matter only simulation of the structure formation and evolution from the beginning to redshift $z=5.5$. The simulation cube is set to be $512$ $h^{-1}$Mpc on a side, to match the transverse scale of SKA strawman survey \citep{McQuinn_2006, 2014MNRAS.439.3262M, 2015aska.confE..10M, Mesinger:2015sha}, and to have a particle resolution $5.6\times 10^6$ \Msun , nearly resolving all dark matter haloes capable of forming galaxy with more than $20$ dark matter particles \citep{Doussot2018SCORCHIR, 2022ApJ...927..186T}. This yields $2$ trillion particles, equivalent to  two state-of-the-art N-body simulations, Euclid Flagship \citep{2017ComAC...4....2P} and Uchuu simulation \citep{2021MNRAS.506.4210I}.  The simulation generates $4$ full particle outputs and $28$ density maps and friends-of-friends (FOF) catalogues between redshift $z=30$ and 5.5, these data will be used for studies of cosmic re-ionisation with post-processing method. 

In this introductory paper, we present the Ultramarine simulation and some basic statistical results about high redshift dark matter haloes, including dark matter halo mass function, halo bias factor and halo mass-concentration relation. The paper is organized as follows. In section 2, we briefly introduce the code to perform the simulation and present details of the simulation. We present our main results in section 3, and give a summary in section 4. 

\section{numerical methods and the simulation}
\label{sec:simu}

\subsection{The Code}

The code to carry out the Ultramarine simulation is PhotoNs-3.4 which is a substantial update from \citet{Wang_2021}. The first version of PhotoNs code was designed to perform massive $N$-body Cosmological simulations on the heterogeneous supercomputer platform \citep{2018RAA....18...62W}. The code adopts a hybrid scheme to compute gravity, with a Particle-Mesh(PM) to calculate the long-range force, a tree method to calculate the short range force and the direct summation Particle-Particle (PP) to calculate interactions from very close particles. As shown in \citet{2018RAA....18...62W}, results from the simulation performed with PhotoNs code are in excellent agreement with that run with Gadget, including power spectrum, the halo mass function as well as internal structure of dark matter haloes. 

later{\color{red},} we replaced the short range gravity calculation of the Photons code with a truncated Fast Multipole Method (FMM),  which has the attracting feature with a time complexity O(N), more suitable for carrying out extreme large simulations. Thus the calculation of gravitational interaction the current code is specifically separated into three parts, long-range PM, FMM operations and PP direct interaction \citep{Wang_2021}. The FMM algorithm contains a series of operations, Particle-to-Multipole (P2M), Multipole-to-Multipole (M2M), Multipole-to-Local (M2L), Local-to-Local (L2L), Local-to-Particle (L2P) and Particle-to-Particle (P2P). The precision and accuracy of such splitting method can be controlled by the tree traversal criteria. One of two key operations, M2L, its $p$-th order operation is computed with the equation 

\begin{equation}
\begin{aligned}
(-1)^{p}r_t^{2p+1}f_{(p)}(x) &= \frac{(2p-1)!!}{x^{{2p+1}}}  {\rm erfc} \left( \frac{x}{2} \right) \\
&+ \frac{1}{\sqrt{\pi} }\sum_{q=1}^{p}  \frac{2^{q-p}(2p-1)!!}{(2p-2q+1)!!}\frac{e^{ -{x^2}/{4} } }{x^{2q}}, 
\end{aligned}
\end{equation}
where ${x \equiv {r}/{r_t}}$ and $r_t$ is the force splitting scale. The other key operation, P2P interaction, is computed by 
\begin{equation}
\label{eq:fp2p}
{\bf f}_{\rm P2P} = -\frac{\bf r}{r^3}\left[ {\rm erfc} \left( \frac{r}{2r_{t}} \right) + \frac{r}{r_{t}\sqrt{\pi}} {\exp}\left(-\frac{r^2}{4r_{t}^2}\right) \right],
\end{equation}
where $r_{t}$ is fixed to be $\sim 1.25$ times grid size and the cutoff radius is $\sim 5.6$ times grid size to control the anisotropic error. Note, the Eq.~\ref{eq:fp2p} is identical to \citet{2002JApA...23..185B}. In practise, the pairwise interaction computed by direct P2P summation is the dominant calculation of our algorithm. Usually, it takes more than $\sim 90 \%$ computing time.  To match such amount of computational demanding, we rewrite the computing kernels on GPU accelerators. Briefly, an interpolation method is employed to compute the exponential truncation function in Eq.~\ref{eq:fp2p}, the interpolating table size is designed to match the L1 cache of cores with sufficient accuracy. The kernel calculation is rearranged at assembling level. We also optimize the task parallelism and  memory access.  We refer the readers to \citet{Wang_2021_gpu} for details.  After those optimizations, the computing efficiency of the kernel calculation is dramatically improved by $\sim 50$. In addition, in order to improve the scalability, in this new version we further improve the communication and imbalance by re-designing different domain decomposition schemes for PM and FMM solver, which are also detailed in Appendix \ref{apx:domain}. 

\begin{figure}
\centering
\includegraphics[width=\linewidth]{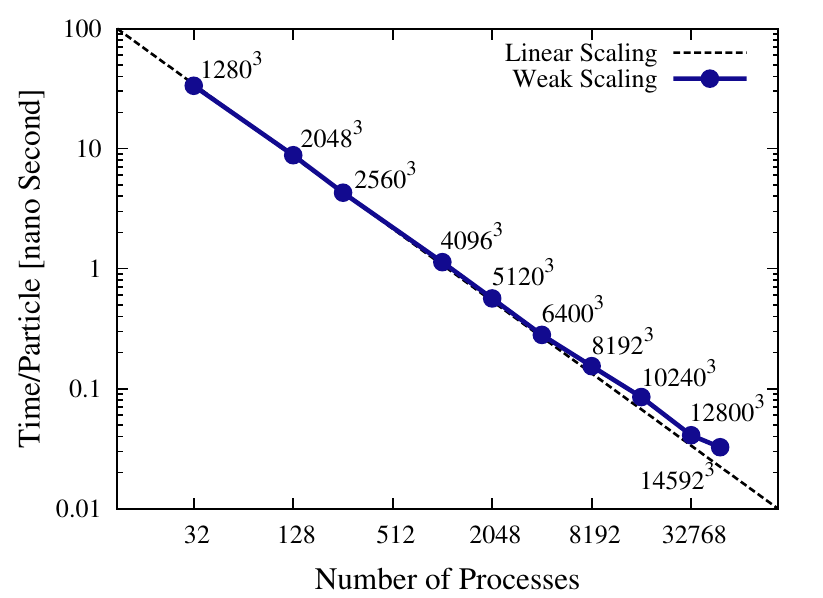}
\caption{Weak scaling performance of the PhotoNs code. The solid symbols show time per particle as a function of the number of computing processors, the total number of particles of each test run is labeled above each data point. The dashed lines show a perfect linear weak scaling relation.}
\label{fig:weak}
\end{figure}

 In Figure~\ref{fig:weak}, we present the weak scaling performance of our code. The test runs were performed with the number of simulated particles varying from  $1280^3$ to $12800^3$ on processes varying from $32$ to $49152$. The vertical axis show wall-clock time for a complete time step calculation normalised by the number of evolved particles. The test results are shown with solid dots, the dashed lines show a perfect linear scaling relation. Clearly the weak scaling of the code is extremely good, with the ideal linear relation for the number of simulated particles from $1280^3$ to $8192^3$. Only when the total number of particles exceed the latter number, the scaling slightly deviates the linear relation, with the weak scalability of $\sim$ 2.1 trillion ($12800^3$) particles achieving 82.51\%, in relative to $\sim$ 2.1 billion ($1280^3$). 

\subsection{The Ultramarine simulation}
The Ultramarine simulation evolves $12800^3$ dark matter particles in a periodical cube of 512 $h^{-1}$Mpc on a side. The number of particles is identical to the state-of-the-art simulations of Ecluid flagship and Uchuu {\citep{2017ComAC...4....2P, 2021MNRAS.506.4210I}}, while the simulation volume of the Ultramarine is a factor of about $200$ times smaller than them, and thus provides $200$ times better mass resolution, which is $5.61\times10^6$ \Msun per particle. The force resolution is set to be $\sim 1.2 ~h^{-1}$ kpc. The Ultramarine simulation assumes Planck Cosmology:  $\Omega_{\mathrm \Lambda}$=0.684, $\Omega_{\mathrm c}$= 0.265, $\Omega_{\mathrm b}$ = 0.0494, H$_0$= 67.32  km$~s^{-1}$Mpc$^{-1}$, $\sigma_{8}$=0.812 and $n_s$ = 0.966. The simulation is carried out from initial redshift $z=99$ to an epoch $z=5.5$, right after reionization, on  ORISE Supercomputer. The initial conditions of the simulation was generated with the traditional Zel'dovich displacement approach   \citep{1985ApJS...57..241E,  2015ascl.soft02003S}, assuming the total matter distribution follows the linear power spectrum calculated by  {\small CAMB} \citep{2000ApJ...538..473L}. Note, using the second order method (2LPT)  ~\citep{2006MNRAS.373..369C} can significantly lower starting redshift, and improve the accuracy of initial conditions at small scales.  \citep{2013MNRAS.431.1866R,2016JCAP...04..047S}. However it require much more memory to generate initial conditions with 2LPT method, limited by the available computing resource, we did not adopt it in the paper.

\begin{figure*}
\centering
\includegraphics[width=0.49\linewidth]{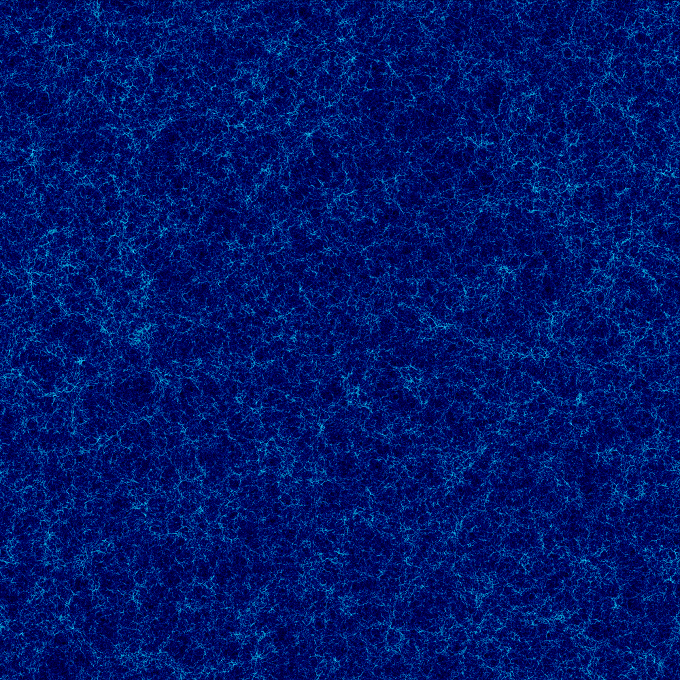}
\includegraphics[width=0.49\linewidth]{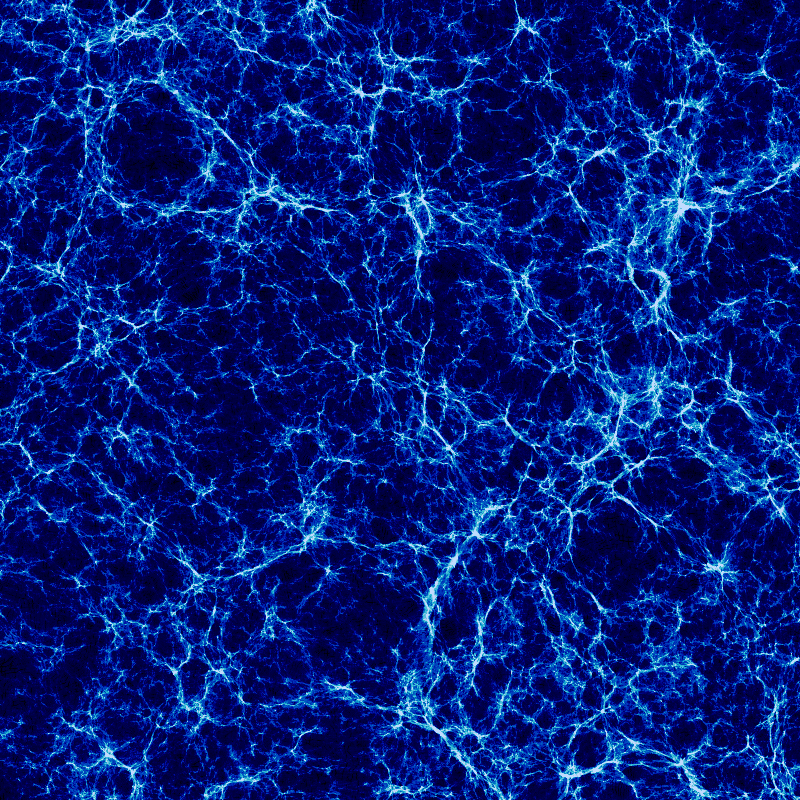}
\caption{Projected dark matter density map at redshift $z=5.5$. The left panel show dark matter distribution in a slice $512$ $h^{-1}$Mpc on a side and 3.2 $h^{-1}$Mpc thick. The right panel show a ''zoom-in'' sub-region of the left with a scale $50$ $h^{-1}$Mpc.} 
\label{fig:density}
\end{figure*}

For a visual impression of matter distribution at high redshift,  in figure~\ref{fig:density} we present a slice of density field at $z=5.5$, with the left panel showing a full scale simulation box $512 ~h^{-1}$Mpc, and the right panel displaying a ``zooming-in'' sub-region of the left with a side-length of $50 ~h^{-1}$Mpc. Interestingly the cosmic web shown in the right panel is qualitatively very similar to what we see at present day. 

Due to huge data size and limited disk storage, we only output particle positions at  $4$ epochs, $z=9.949, 7.667, 5.985$ and $5.5$. The particle position data is compressed with an algorithm described in Appendix \ref{apx:cmpr}.  In addition, we output density maps with $6400^3$ regular meshes for $28$ epochs, these maps are useful for future re-ionisation studies with post-processing. Dark matter haloes containing a minimal $20$ particles are identified with a build-in on-the-fly Friends-of-Friends (FOF) halo finder. Note, the mass of a dark matter halo in this paper is defined as its FOF mass unless otherwise stated. The redshift of the outputs is listed in in Table ~\ref{tab:outputlist}. 

\begin{table}
\centering
\begin{tabular}{rrrrrrrr}
\hline
\multicolumn{2}{r}{redshift} & \multicolumn{2}{r}{redshift} & \multicolumn{2}{r}{redshift} & \multicolumn{2}{r}{redshift} \\
\cmidrule(r){1-2} \cmidrule(r){3-4} \cmidrule(r){5-6} \cmidrule(r){7-8}
 1 & 29.5179 &  8 & 15.2634 & 15 &  9.1896 &  22 &  6.9217 \\
 2 & 26.8937 &  9 & 13.8649 & 16 &  8.8296 &  23 &  6.6419 \\ 
 3 & 24.4951 & 10 & 12.5867 & 17 &  8.4824 &  24 &  6.3719 \\ 
 4 & 22.3028 & 11 & 11.4184 & 18 &  8.1417 &  25 &  6.1115 \\
 5 & 20.2990 & 12 & 10.3505 & 19 &  7.8242 &  26 &  5.8602 \\
 6 & 18.4675 & 13 &  9.9496 & 20 &  7.5125 &  27 &  5.6179 \\
 7 & 16.7935 & 14 &  9.5628 & 21 &  7.2118 &  28 &  5.5000 \\ 
\hline
\end{tabular}
\caption{Output list of the density maps and halo catalogues at 28 epochs in the range of $z \in (30,5.5)$.}
\label{tab:outputlist}
\end{table}

\begin{figure}
\centering
\includegraphics[width=1.0\linewidth]{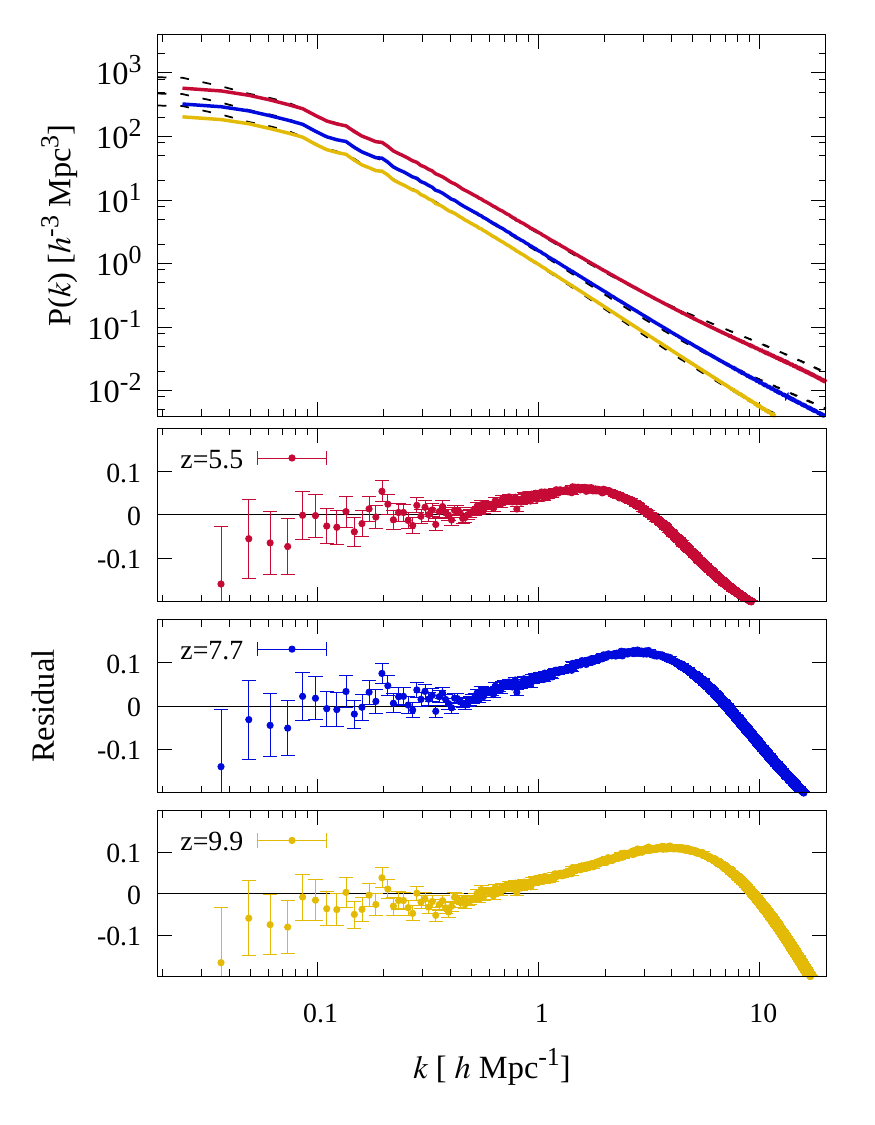}
\caption{Mass power spectrum of the simulation at $z=5.5, 7.7, 9.9$ (top to bottom) in the top panel. The dashed lines show predictions from the halo fit model. The bottom panels show the residuals of power spectrum with respect to the halo fit.} 
\label{fig:ps}
\end{figure}

The power spectra are on-the-fly calculated on 28 redshifts, from 93.8 to 5.5, and we show three of them at $z=9.9, 7,7, 5.5$ in Fig.~\ref{fig:ps}, where the dashed curves are predictions of {\small CAMB} with the halo-fit model \citep{2000ApJ...538..473L, 2015MNRAS.454.1958M} and the solid lines are the measured power spectra. In the linear regime, the power spectrum is consistent with the theoretical growth. Note, we estimate power spectrum with a $12800^3$ Cloud-in-Cell (CIC) mesh. The numerical suppression at non-linear region ( at high $k \ge 10 $) is a well-known effect (e.g. \citet{2005ApJ...620..559J}). Apparent differences between our simulation and the halo fits in the wave number range $1<k<10$ can be observed, while at this regime the model is not well tested.  The wall-clock time for completing the simulation is about $14.6$ hours, and the on-the-fly post-processing takes about $\sim 2$ more hours.

\section{Properties of dark matter haloes at high z}
\label{sec:resu}

In this section, we take advantage of the high resolution and large volume of the Ultramarine simulation to analyses some basic statistics of dark matter haloes at high redshifts, including halo mass functions, linear halo bias factor and the halo mass-concentration relation. 

\subsection{{The FOF halo Mass Function}}
The abundance of haloes provides important information on the abundance of galaxies during the re-ionization. Extensive effort has been made to build analytical models and calibrate them with N-body simulations. However, most models were calibrated with low redshift relative massive haloes except of very limited works focusing on the halo mass function at high redshift. Here we extend these studies to a halo mass down to about $10 ^{8}$ \Msun for redshifts ranging from $z=5.5$ to $z=9.9$, and compare our numerical results with three representative models. The models include an analytical one based on elliptical collapsing theory\citep{2001MNRAS.323....1S}, a fitting model to predict the halo abundance across dark age \citep{2007MNRAS.374....2R} and the other fitting model calibrated to match relatively massive haloes at low redshift.

\begin{figure*}
\centering
\includegraphics[width=0.45\linewidth]{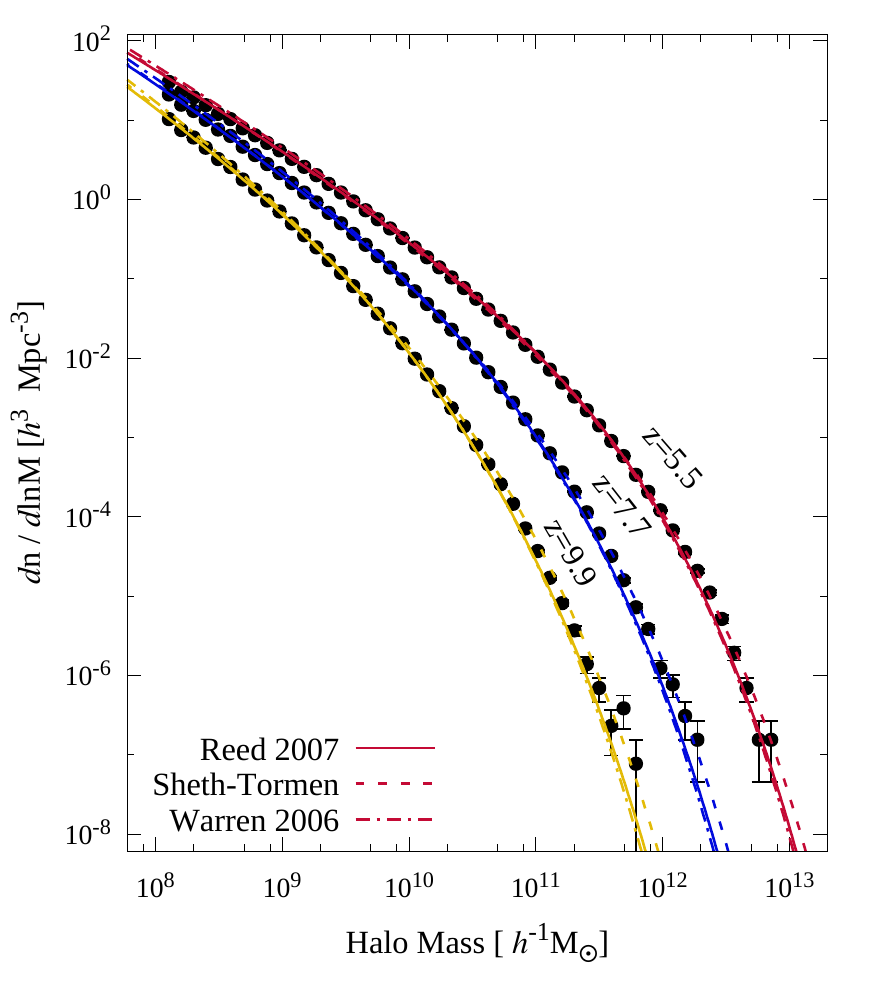}
\includegraphics[width=0.45\linewidth]{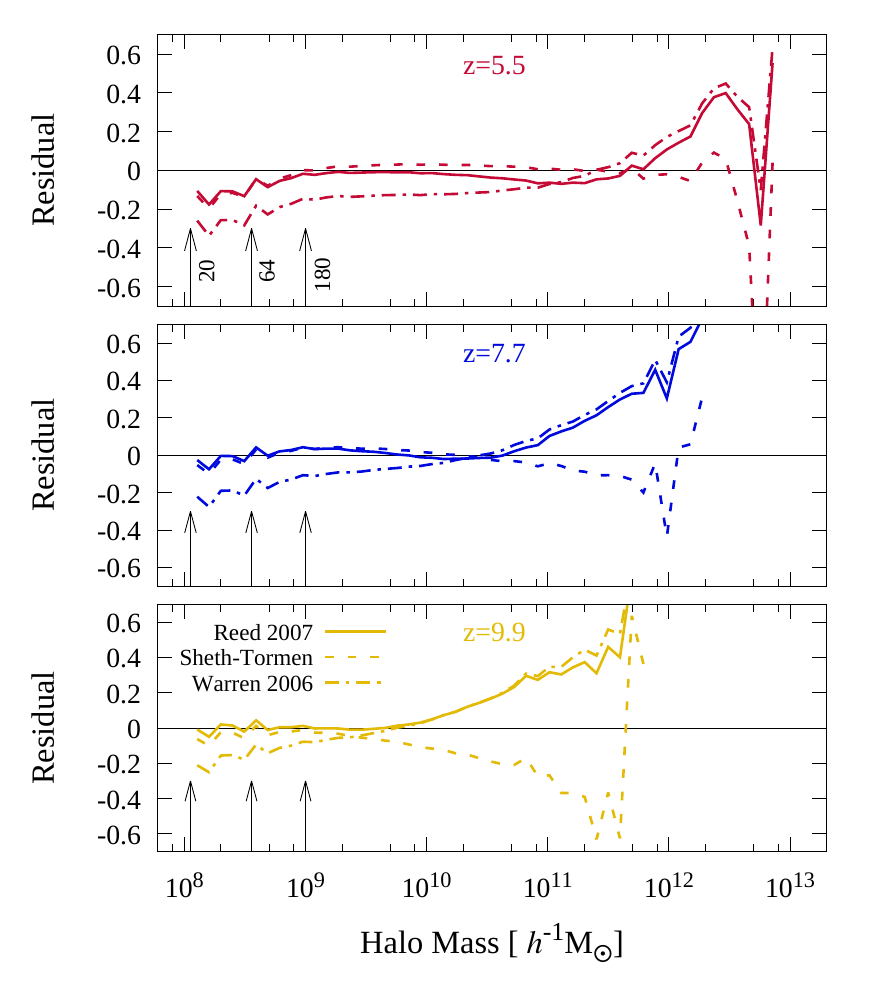}
\caption{Comparison of halo mass function in the Ultramarine simulation with some analytic predictions. The solid points show our numerical results, the solid, dashed and dashed-dotted lines show prediction from the model of \citet{2007MNRAS.374....2R}, \citep{ST1999} and \citet{2006ApJ...646..881W}, respectively. The right panel shows the residuals of our numerical results in relative to the three models $\delta=$ data/model-1. The arrows in the right panel show that halo mass containing $20$, $64$ and $180$ particles.} 
\label{fig:mf}
\end{figure*}

The comparison is presented in Figure~\ref{fig:mf}, the numerical results from our Ultramarine are shown with solid dots, and the theoretical predictions from  \citet{2007MNRAS.374....2R}, \citet{ST1999}  and \citet{2006ApJ...646..881W}  are shown with the solid, dashed and dashed-dots lines, respectively. Results for different redshifts are distinguished with different colors as indicated in the label.  In order to emphasis the differences, we plot the residuals between the models and simulation in the right panel of the same plot. Arrows indicate haloes containing 160, 64 and 20 particles, respectively. Overall all three models provide reasonable predictions on the halo mass functions at three epochs, $z=5.5$, $7.7$ and $z=9.9$, while the degree of the agreement among the models varies with redshift and halo mass. At relative high mass end, $>6 \times 10^{11}$\Msun at $z=5.5$,  $>4 \times 10^{10}$\Msun at $z=7.7$ and  $10^{10}$\Msun at $z=9.9$,  \citet{2007MNRAS.374....2R} and \citet{2006ApJ...646..881W} over-predict the halo mass function by up to 50 percents, while the original Sheth-Tormen formulae is in well agreement with our numerical data except of very high mass end, in that regime our numerical data is noisy.  At low mass end,  \citet{2007MNRAS.374....2R}  and Sheth-Tormen models quite accurately agree with our numerical data, while the fitting model of \citet{2006ApJ...646..881W}  underestimates the halo function by up to 20 percent. 

\subsection{FOF halo Bias}
Halo bias describes the distribution of dark matter haloes relative to the underlying mass density field, and thus is instrumental for many theoretical applications, for instance, populating galaxies with halo occupation distribution model (HOD). According to pioneering analytical works based on extended Press-Schester theory, for example, \citet{1996MNRAS.282..347M}, the two-point correlation functions of the dark matter halo of mass $M$, $\xi_{hh}$  approximately parallel to that of underlying dark matter with bias factor $b$, 
\begin{equation}
\xi_{hh}(z) = b^2 \xi_{mm} (z).
\end{equation}

In the peak-background split (PBS) framework of  \citet{1989MNRAS.237.1127C, 1996MNRAS.282..347M, ST1999}, the bias factor $b (\nu) = 1+(\nu^2-1)/\delta_c$ .  Here $\delta_c$ is the critical over-density at collapse, $\nu=\delta_c/\sigma(M)$ is the peak height, $\sigma(M)$ is the rms linear mass fluctuation extrapolated to redshift $z$.  The model has been shown to be qualitatively consistent with numerical simulations, yet deviate quantitatively  \citep{1998ApJ...503L...9J, 1998Natur.392..359G, 2000MNRAS.319..209C, 2005MNRAS.362.1451M, 2005MNRAS.363L..66G, 2008MNRAS.387..921A, 2010ApJ...708..469F, 2011ApJ...732..122B, 2016JCAP...02..018L, 2017JCAP...03..059L, 2021MNRAS.507.3412C}. Based on high resolutions simulations, many improved models have been proposed. Again, previous studies in this subject have been focused on relatively massive haloes and at lower redshifts. In this section we exploit the unprecedented dynamical range and statistics of the Ultramarine simulation to extend these studies to high redshifts and to lower halo masses.  

\begin{figure}
\centering
\includegraphics[width=\linewidth]{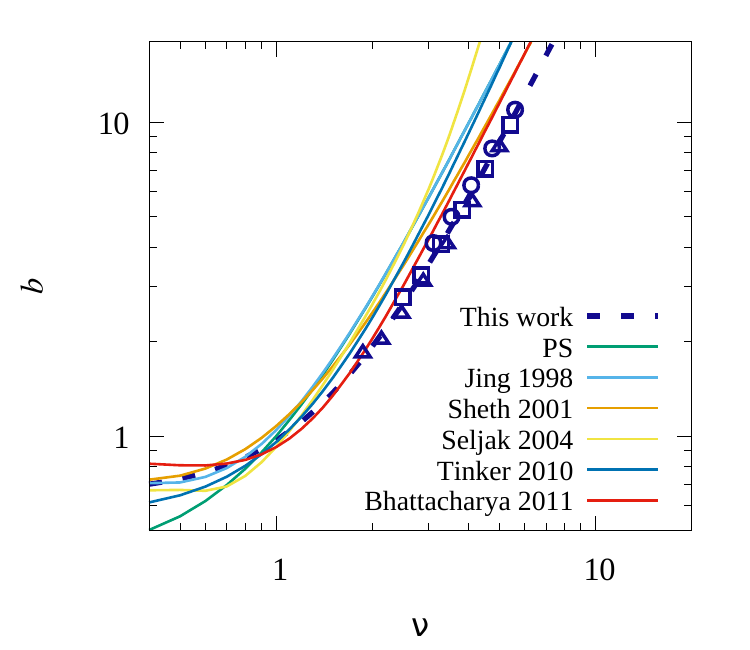}
\caption{Halo bias as a function of peak height $\nu = \delta_c / \sigma(M,z)$. Different symbols show results measured for different redshift as indicated in the label. The different lines refer to different analytic predictions from $6$ models as indicated.}
\label{fig:bias}
\end{figure}

In practise, we use Landy-Szlay method to estimate two point correlation functions for  FOF dark matter halo \footnote{ Most existing models on the halo bias are based on FOF halo} and underlying density field,  $\hat{\xi}_{ii} = (DD - 2DR + RR)/ RR$ , where $DD$ is the averaged pairs $i$ component and $RR$ is the averaged pairs of the random sample \citep{1993ApJ...412...64L}.  The bias factor $b$ is then estimated by following the procedure of \citet{2005MNRAS.363L..66G}, namely minimizing the mean square difference in $\log \xi$  for six bins ranging from  $5.5 < r <25 h^{-1}$kpc. In Figure~\ref{fig:bias}, we present halo bias as a function of  `peak height' $\nu(M_h,z)$ for three epochs, $z=5.5$  (open triangles) , $z=7.7$ (squares) and $z=9.9$. We overplot 6 analytical or fitting models computed with COLOSSUS package ~\citep{2018ApJS..239...35D} for comparison, each model is plotted with different colours as indicated in the label. Note, while being low mass haloes, they are still quite rare objects at the redshifts in terms with peak height, as a result our data only cover large $\nu$ end. Apparently halo bias factors seem only a function of $\nu$ and are independent of redshift, in qualitatively support of the model from EPS theory, while being quantitatively different. At $\nu=3$, EPS model overpredicts the bias factor by about $70$ percent, some models agree with our numerical simulation data better than EPS, but are still much larger than our data at high $\nu$ end.  The dashed lines show a fit to our data with a simple formula of
\begin{equation}
b(\nu) = 1 + \frac{\tilde{a}  (\nu^2 - \tilde{u} ) }{\delta_c}.
\end{equation}
with two free parameters of $\tilde{a} =0.54$ and $\tilde{u}=1.1$. 
Note that the range of $\nu$ in our simulation is $[2, 5.3]$,  the fits may not be valid for small values of $\nu$. In addition, the fits are based on one realization and thus may have some uncertainties.

\subsection{Mass-Concentration Relation}
Previous studies on the structure of dark matter haloes has established that the density profiles of dark matter haloes can be well described with a universal from, Navarro–Frenk–White (NFW) profile \citep{1997ApJ...490..493N},
\begin{equation}
\rho(r) = \frac{\rho_s}{({r}/{r_s})\left(1+{r}/{r_s}\right)^2}.
\label{eq:nfw}
\end{equation}
Here $\rho_s$ is a characteristic density,  and $r_s$ is the scale radius at which the logarithmic slope $\gamma \equiv d\ln \rho/d \ln r=-2$. As shown in the original NFW paper and confirmed by later studies  \citep{2001MNRAS.321..559B, 2002ApJ...568...52W, 2007MNRAS.378...55M, 2014MNRAS.441.3359D, 2014MNRAS.441..378L, 2015MNRAS.452.1217C, 2016MNRAS.460.1214L}, the concentration parameter of dark matter haloes, $c={r}_{200}/r_s$, statically correlate with their masses, with concentration parameter decreasing with increasing halo mass. Here $c={r}_{200}/r_s$, $r_{200}$  is the viral radius  within which the enclosed mass is  200 times critical value.  It is also commonly agreed that the halo mass concentration relation evolves with redshift. There exists some theoretical or fitting models on the halo mass concentration relation, these models are mainly based on simulated results at low redshift $z<3$, and so have been well tested with low redshifts data. It is interesting to test these models at higher redshifts with our Ultramarine simulation which has most powerful statistics at these epochs.

\begin{figure}
\centering
\includegraphics[width=0.97\linewidth]{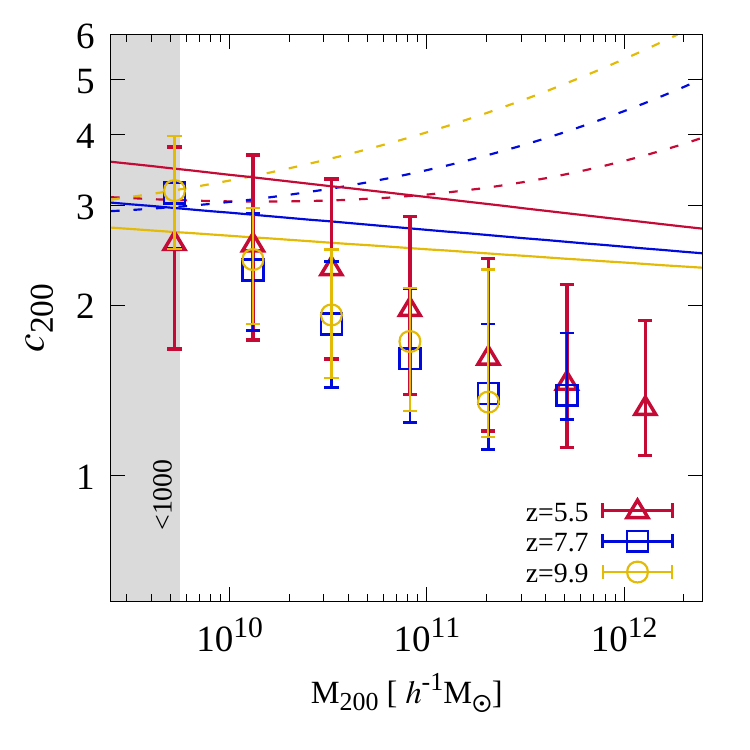}
\caption{Halo mass-concentration relation. Individual symbols show median values of concentration, different symbols refer to different redshift as indicated in the label and the error bars display 1-$\sigma$ scatters. The solid lines show predictions from  \citet{2016MNRAS.460.1214L}, and the dashed lines are the fitting model given by \citet{2019ApJ...871..168D},  predictions for different redshift are distinguished with different color as indicated.  The shaded area show results for the haloes with less than $1000$ particles inside their virial radius. } 
\label{fig:mcr}
\end{figure}

In Figure~\ref{fig:mcr}, we plot the concentration parameter as a function of halo mass for 3 different epochs, $z=5.5, 7.7$ and $9.9$.  Note, the halo mass adopted here is the viral mass $M_{200}$. As the distribution of concentration is quite broad \citep{2007MNRAS.381.1450N,2021A&A...652A.155S}, it is nontrivial to derive mean of concentration, following literature, we show median values of concentration. Results for different redshift are distinguished with different symbols, and error bars indicate $1\sigma$ scatters. The concentration parameter of a dark matter halo is estimated by following the procedure of ~\citet{2008MNRAS.387..536G}, namely we calculate a spherically-averaged density profile by binning the halo mass in equally spaced logarithm bins, between $r_{200}$ and $0.1 ~{r}_{200}$. Then we calculate the concentration of each profile by minimising the rms deviation between $\rho(r)$ and the NFW prediction. We consider only haloes with more than $750$ particles within their viral radius. The shaded area indicates dark matter haloes with less than $1000$ particles.  Numerical convergence studies (e. g., ~\citet{2007MNRAS.376..215B}, \citet{2007MNRAS.381.1450N}) suggest that it requires at least $1000$ particles to faithfully estimate concentration parameter of a dark matter halo. In the same plot, we also show predictions from 2 representative  models of \citet{2016MNRAS.460.1214L} and \citet{2019ApJ...871..168D}. The Ludlow model overall reasonably matches the halo mass concentration relation of dark matter haloes across $20$ orders of magnitude in halo mass\citep{2020Natur.585...39W}, while \citet{2019ApJ...871..168D} model interestingly predicts an upturn trend on the relation at the high mass end, which has been confirmed by few studies/simulations \citep{2015PASJ...67...61I, 2016MNRAS.457.4340K, 2021MNRAS.506.4210I}.

The first noticeable feature of the plot is that the halo mass concentration is still monotonic at high redshifts, with the concentration decreasing with increasing halo mass. This in agreement with most analytical/fitting models, while the measured concentration parameters are much lower than the model predictions. Only at the lowest mass end, our data marginally agrees with Ludlow et al. Our data thus doesn't support the upturn feature reported by \citet{2012MNRAS.423.3018P, 2015ApJ...799..108D, 2019ApJ...871..168D}, and the \citet{2009ApJ...707..354Z} model which predicts that high redshift haloes all have same concentration parameters about $4$. The second noticeable feature is that the halo mass concentration relation has little evolution from $z=10$ to $z=5.5$, in contrast to strong evolution predicted by \citet{2016MNRAS.460.1214L}.  

To examine whether high redshift dark matter haloes can be described with NFW model, in Figure~8 we show stacked density profiles of dark matter haloes in a narrow mass range $[3.0, 3.3]$ $\times 10^{9}$ \Msun and their best NFW fits. Again, results for $3$ different epochs are presented. Note the density profiles are shown with $\rho(r) r^2$ in order to remove dominant radial dependence, and the profiles for different redshifts are arbitrarily normalized to make results distinguishable. Comparing density profiles (solid symbols) and their best NFW fits (solid lines) shown in bottom of the plot, NFW formulae provides reasonable fits to high redshift dark matter halo profiles, especially for redshift $z=5.5$ and $z=7.7$. For the two epochs only the out-most points are slightly smaller than the best fitted values. NFW fits are slightly worsen for $z=9.9$, but are still acceptable in the most radial range. Results shown in the upper plot for relaxed halo are similar. Here we follow \citet{2007MNRAS.381.1450N} to select relaxed haloes but only use center-offset criteria $s= | r_{\rm c}-r_{\rm cm}| \le 0.07 \sim {r}_{200}$, here $r_{\rm c}$ is the potential centre and $r_{\rm cm}$ is the barycenter.  Note \citet{2007MNRAS.381.1450N} adopted $2$ more criteria to judge the equilibrium state of a halo, namely substructure fraction and virial ratio, but we currently neither identify substructure nor have output velocity information, and hence only adopt the centre-offset criteria.

\begin{figure}
\centering
\includegraphics[width=\linewidth]{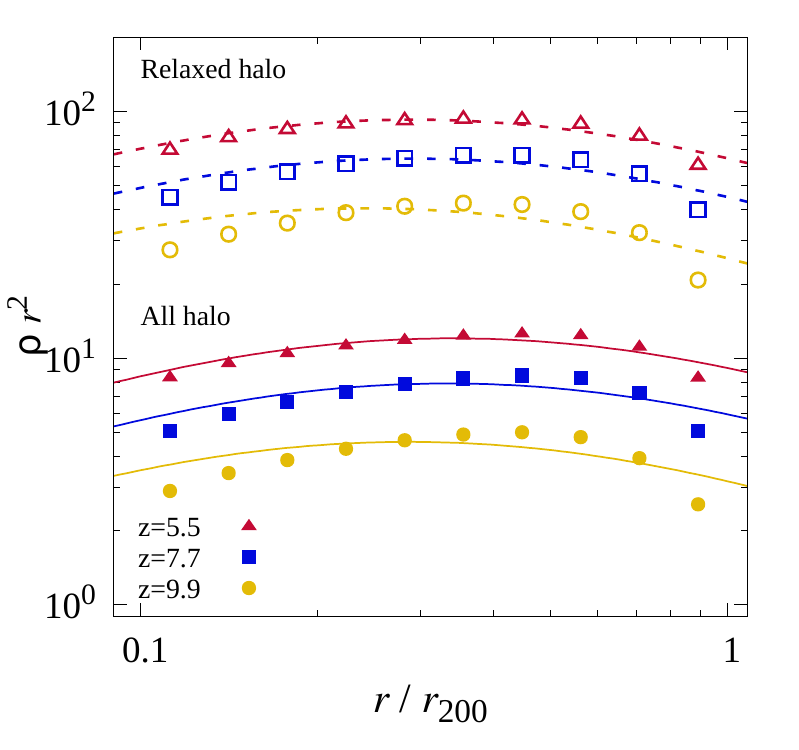}
\caption{The stacked density profiles of dark matter haloes of mass $3.3 \times 10^9 h^{-1}$M$_{\odot}$ at $3$ epochs, and the lines show their corresponding best NFW fittings.  The density profiles are plotted by multiplying $r^2$ and arbitrarily normalized, results for all and only relaxed are shown in the lower and upper part of the panel, respectively. }
\label{fig:stack}
\end{figure}

\section{Summary}

We performed an extremely large simulation, the Ultramarine simulation,  to resolve structure formation and evolution from beginning to redshift just after cosmic re-ionization. The code to carry out the simulation is PhotoNs, which fully takes advantage of computing power of acceleration card and thus is quite efficient for extremely large $N-$body simulation. The wall-clock time to complete the $2$ trillion particles Ultramarine simulation is only $14.6$ hours. 

In this introductory paper, we present some basic results of the simulation, including the halo mass function, halo bias and halo mass-concentration relation. Comparing with existing models, either Sheth \& Tormen or Reed et al. model describes high redshift halo mass function of our simulation results quite well, while none of existing models for the halo bias or halo mass concentration relation match our numerical data. All halo bias models compared in this paper overestimate high redshift halo bias by large factors, we give a simple fit to our numerical data. We find that high redshift dark matter haloes can be reasonably described by NFW model, while concentration parameters of them are well below predictions from the models quoted in the paper. In particularly, we do not see the up-turn feature at high mass end of the halo mass concentration relation as firstly reported by \citet{2012MNRAS.423.3018P} and recently confirmed by \citet{2019ApJ...871..168D, 2021MNRAS.506.4210I}

The Ultramarine simulation has a mass resolution $5.6\times 10^6$ \Msun, marginally resolve all dark matter haloes exceeding $\sim 10^8$ \Msun. Twenty-eight density fields as well as dark matter halo are quite useful to model re-ionisation process with post-processing technique, which will be presented in our future studies.

\section*{Acknowledgements}
We acknowledge th e support from National SKA Program of China (Grant No. 2020SKA0110401),  National Natural Science Foundation of China (Grant No. 11988101, 12033008) and K.C.Wong Education Foundation. The simulation is carried out on ORISE Supercomputer, China. WQ thanks the advice about the epoch of reionization from Y. Xu, B. Yue and the parallel strategy by the useful discussions with J. Makino and M. Iwasawa.

\section*{DATA AVAILABILITY}
The data underlying this article will be made publicly available online after publication of the article.

\bibliographystyle{mnras}

\appendix
\section{Domain decomposition}
\label{apx:domain}

\begin{figure}
\centering
\includegraphics[width=\linewidth]{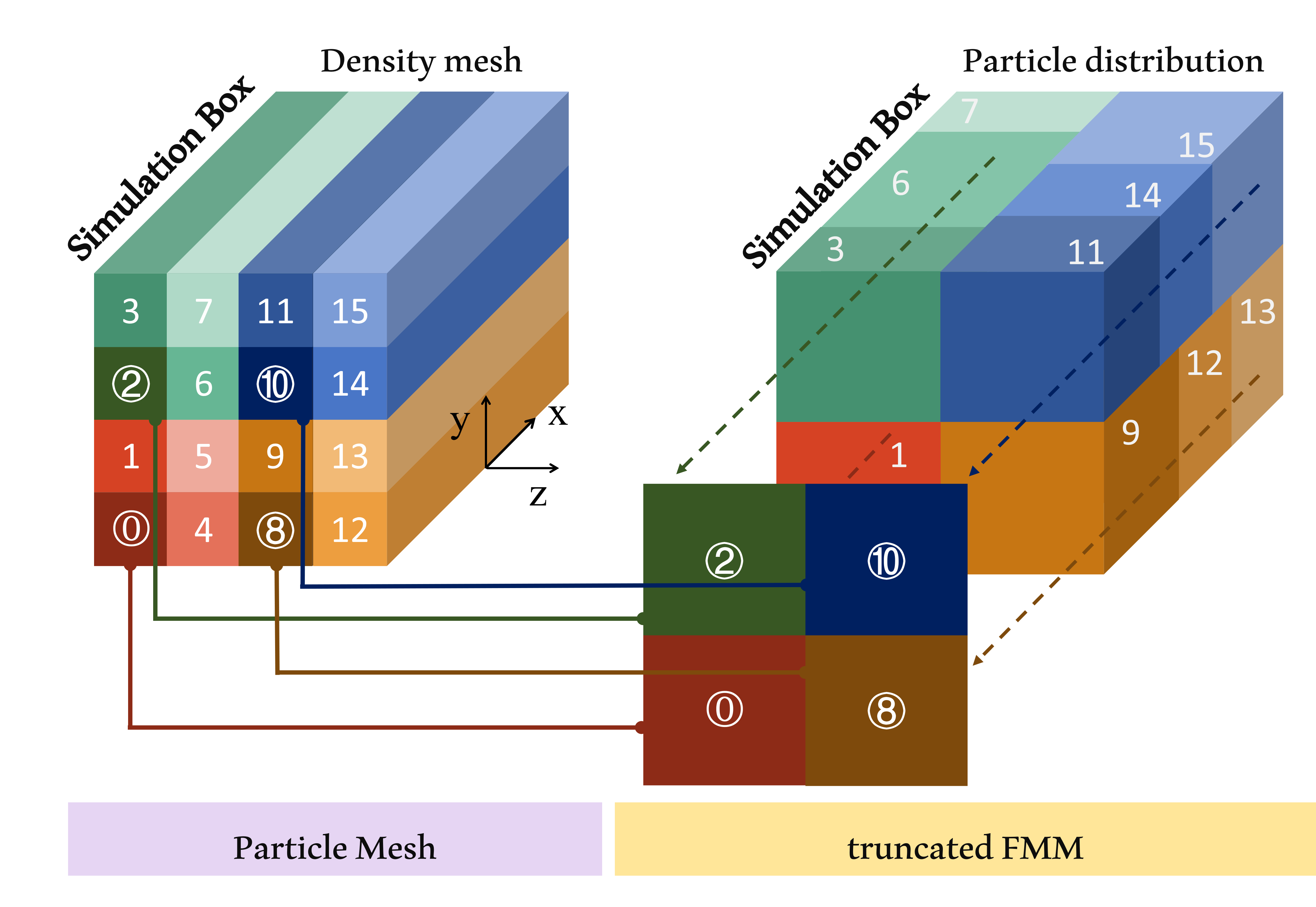}
\caption{Domain decomposition. The left box represents the decomposition of particle-mesh solver and right box represents the decomposition to compute truncated FMM for the short-range gravity. }
\label{fig:domain}
\end{figure}

In this work, we adopt a new decomposition approach and two-layer communicators to improve the scalability of Photons code. We bind several adjoin computing domains into a single group. Thus the peer-to-peer communication between any domains is implemented by a two layer communicators. First, the information of computing domain boundaries is gathered into a head process via the intra-communicator, then the collection of boundary information is transported to the destination domain via the higher level communicator (head layer). The computing domains, excluding head domains (similar with the super domain in \citet{2016PASJ...68...54I}) and \citet{2009PASJ...61.1319I}, are in charge of FMM calculation, but all processes are involved into the PM convolution based on FFT. There exists a special direction along x-axies in the simulation box, due to the interface of FFT library, 2DECOMP\&FFT \citep{20102decomp}. Hence the domain binding is also along with $x$-axies as shown in Fig~\ref{fig:domain}. 

For instance, we consider a toy run carried out by 16 processes. The upper right 4 domains contain rank 10, 11, 14 and 15 are bound together as the blue group and rank 2, 3, 6 and 7 are bound as the green group. The rank 2 and 10 are employed as head domains and rank 11, 14, 15, 3, 6, and 7 are employed for FMM calculation. The head layer communicator consists of rank 0, 2, 8 and 10.  Those 4 blue domains (the right box) exactly locate at the same position for their FFT mesh configuration for PM solver (the left box). The PM mesh constructed from rank 11, 14 and 15 is still suit for the FFT configuration on the same ranks. Thus the amount of communication is more regular and smaller than the previous version. 

On the other hand, the particle distribution is inhomogeneous in one group. In each group, the inner boundaries are separately determined by the particle distribution. At each step, we measure the number of particles along the $x$-direction to compute the coordinates of inner boundaries of group. This method also works for the balance of work-load by measuring the computing task counts.  

\section{The compressed algorithm}
\label{apx:cmpr}

In our code implements, all particles are organized in a series of leaves (the finest tree nodes) and  each leave is a compact box containing hundreds of adjoining particles. According to IEEE standard 754, the sign, exponent and mantissa of a floating point (FP32) contain 1, 8, and 23 bits, respectively. First, we normalize the particle coordinate into the unit so that the coordinates in the same leaf always share the same sign, exponent and partial mantissa bits due to the locality. Thus, the same bits in a pack can be extracted as the origin coordinate, the deviated bits are considered as an offset with respect to the origin. Therefore, only deviated offset bits need to be recorded for each particle and the origin once for a leave pack. With such a compressed implementation, the relative error can be controlled under $10^{-6}$ by using 20-bit precision of mantissa. For the 20-bit version, the compression ratio is about 2, for the uniformed distribution of particles. In contrast, the $gzip$ gives the compression ratio is about 1.6, and the $xz$ is about 1.8 to a primitive simulation snapshot. Practically, we use 9-12 bits to record the offset. Despite such procedure will lost information, the compression ratio can be improved up to about 2.6 $\sim$ 3.4.

\bsp
\label{lastpage}
\end{document}